\documentclass[reprint,amssymb,aip,longbibliography]{revtex4-2}
\usepackage[utf8]{inputenc}
\usepackage[T1]{fontenc}
\usepackage{graphicx}
\usepackage{amsmath}
\usepackage{amssymb}
\usepackage{mathrsfs}

\usepackage{braket}
\usepackage{geometry}
\usepackage{here}
\usepackage{lmodern}
\usepackage{array}
\usepackage{natbib}
\usepackage{url}
\usepackage{textcase}
\usepackage{bm}
\usepackage{natbib}
\usepackage{color}
\begin{document}

\title{Interference is in the eye of the beholder: application to the coherent control of collisional
processes}

	
	\author{Adrien Devolder$^{1}$, Timur V. Tscherbul$^{2}$, and {Paul Brumer$^{1}$}}
	
	\affiliation{$^{1}$Chemical Physics Theory Group, Department of Chemistry, and Center for Quantum Information and Quantum Control, University of Toronto, Toronto, Ontario, M5S 3H6, Canada\\
		$^{2}$Department of Physics, University of Nevada, Reno, NV, 89557, USA}
	
	\begin{abstract}
	
Interference is widely regarded as a foundational attribute of quantum mechanics. However,
for a given experimental arrangement, interference can either contribute or not contribute 
to the outcome depending upon the basis in which it is measured. This observation is both
foundational and particularly relevant to coherent control of molecular processes, an approach
based upon quantum interference. Here we address this issue  and its relevance to controlling 
molecular processes via the ``coherent control scattering (CCS) matrix", a formalism that allows an analysis  of modifications in interference structure resulting from a change of basis. This analysis reveals that the change in interference structure can be attributed to the non-commutativity of the transformation matrix with the  CCS matrix, and the non-orthogonality of the transformation. Additionally,
minimal interference is shown to be associated with the CCS eigenbasis, and that
the Fourier transform of the eigenvectors of the CCS matrix provides the maximal interference and hence the best coherent control. The change of controllability through a change of basis is illustrated with an example of $^{85}$Rb+ $^{85}$Rb scattering. In addition, the developed formalism is applied to explain recent experimental results on He + D$_2$ inelastic scattering demonstrating the
presence or absence of interference depending on the basis.

	\end{abstract}
	\date{\today}
	\maketitle
	\section{Introduction}

Interference lies at the heart of quantum mechanics. In particular,
when two or more indistinguishable pathways to the same final state exist, the final state amplitudes can cancel each other or enhance 
the probability of observation beyond the sum of the
probabilities of the independent paths. Most interesting, although not 
widely emphasized, is that interference in a given physical arrangement 
depends upon the basis in which it is analyzed and/or experimentally
measured. Thus, the outcome in a given experiment may be interpretable as a result 
of interference between independent contributions or as arising
from interference-independent routes. For example, a reinterpretation of a recent He + D$_2$  inelastic scattering 
experiment \cite{Zhou2021}, discussed in Sec. VI below,  shows that 
it can either be viewed as displaying interference or not depending on the basis in which
it is viewed. It  will serve as a specific example of the analysis discussed below.

The methods introduced here are generally applicable, but  
our focus is on the basis set dependence of interference 
within the framework of the coherent control of collision phenomena. 
Coherent control\cite{Brumer_book,Shapiro1996} is a quantum approach to controlling 
molecular processes based upon interference phenomena, making this issue of the basis set
dependence of interference particularly relevant. To be precise, we consider a fixed physical scattering arrangement viewed from
different bases and show that interference can be relevant in one basis but 
not in another. Therefore, the controllability by tuning the relative phases in the initial superposition can change between two different basis. Although we choose to focus here on scattering,
the concepts introduced are applicable to a wider variety of quantum processes. 

This article addresses these concerns using a matrix-based approach to interference and
coherent control, originally introduced by Frishmann et al. \cite{Frishman1999} for the optimal control problem. This matrix is referred to as the coherent control scattering (CCS) matrix throughout the paper. In section II, we show how coherent control can be expressed using the CCS matrix and in section III how the controllability can be analyzed with this matrix. In section IV, we illustrate how the CCS matrix formalism naturally accommodates changes of basis, and establish the conditions under which a change of basis alters the interference structure, possibly disappearing in the new basis. In particular, we discuss the diagonalisation of the CCS matrix, which leads to a non-interference scenario and  by contrast, we demonstrate how the quantum Fourier transforms can create a basis where interference is maximized. Section V illustrates the consequences of the change of basis on the controllability with the example of ultracold $^{85}$Rb+ $^{85}$Rb ultracold scattering. Section VI applies this formalism to a recent experiment on He + D$_2$ scattering to 
demonstrate the appearance  interference (or lack thereof) in the interpretation of the observed differential cross section. We summarize in section VII. 

	\section{Coherent control matrix}
Consider a scattering process between two molecules or atoms, prepared in a superposition of $N_i$ internal states from the basis $\left\{\ket{i}\right\}$:
	\begin{equation}
		 \ket{\Psi_{ini}}=\sum_{i=1}^{N_i} a_i \ket{i}. 
		 \label{eq:ini}
	\end{equation}
The differential cross section for the transition to a set of $N_{j'}$ final internal states $\{\ket{j'}\}$  can be expressed as:
	\begin{equation}
		\frac{d^2\sigma}{d\theta d\phi}(\theta,\phi)=\sum_{f_{int}=1}^{N_{j'}} \left|\sum_{i=1}^{N_i}  a_i f_{i\rightarrow j'}(\theta,\phi)\right|^2,
		\label{eq:diff}
	\end{equation}
where $f_{i\rightarrow j'}(\theta,\phi)$ is the scattering amplitude from the initial state $\ket{i}$ to the final internal state $\ket{j'}$.  The number of final internal states can vary depending on the specific phenomena being studied. For example, if state-to-state transitions are considered, there will be only one final internal state, whereas if the focus is on a specific reaction, the number of final internal states will equal the number of channels for that reaction.
	
The coherent control scattering matrix (CCS)
	\begin{equation}
		\mathcal{C}_{ij}(\theta,\phi)=\sum_{j'=1}^{N_{j'}}f_{i\rightarrow j'}(\theta,\phi)f_{j\rightarrow j'}^*(\theta,\phi)
		\label{eq:CCSM}
	\end{equation}
	can be used to rewrite Eq. (\ref{eq:diff}) as a Rayleigh quotient:
	\begin{equation}
		\frac{d^2\sigma}{d\theta d\phi}=\boldsymbol{a}^\dagger \boldsymbol{\mathcal{C}}\boldsymbol{a}.
		\label{eq:diff2}
	\end{equation}
Here, $\boldsymbol{a}$ is a unit row vector $\begin{pmatrix} 
		a_1 & a_2 & ... & a_N
\end{pmatrix}$ containing the superposition coefficients. Since $\boldsymbol{a}^\dagger\boldsymbol{a}=1$, the denominator of the Rayleigh quotient is omitted here or in the rest of the paper.
This formulation of coherent control as a Rayleigh quotient using the CCS matrix can be extended to the $\phi$-independent (half-integrated) differential cross section  after integration over the azimutal angle $\phi$:
	\begin{equation}
		\frac{d\sigma}{d\theta}=\boldsymbol{a}^\dagger \boldsymbol{\mathcal{C}}\boldsymbol{a}
		\label{eq:half}
	\end{equation}
	with 
	\begin{equation}
		\mathcal{C}_{ij}(\theta)=\sum_{j'=1}^{N_{j'}}\int d\phi f_{i\rightarrow j'}(\theta,\phi)f_{j\rightarrow j'}^*(\theta,\phi)
	\end{equation}
	and to the integral cross section
	\begin{equation}
		\sigma=\boldsymbol{a}^\dagger \boldsymbol{\mathcal{C}}\boldsymbol{a}
		\label{eq:int}
	\end{equation}
	with 
	\begin{equation}
		\mathcal{C}_{ij}=\frac{\pi}{k^2}\sum_{j'=1}^{N_{j'}}\sum_\ell \sum_{\ell'} T_{i,\ell\rightarrow j',\ell'}T_{j,\ell\rightarrow j',\ell'}^*.
		\label{eq:F_int}
	\end{equation}
Here $T_{i,\ell\rightarrow j',\ell'}$ is the $T$-matrix between the initial state $\ket{i}$ in the initial partial wave $\ell$  and the final state $\ket{j'}$ in the final partial wave $\ell'$, and $k$ is the relative initial momentum. Equation (\ref{eq:F_int}) was derived using the definition of scattering amplitudes and the orthogonality of the spherical harmonics \cite{Book_Krems}. When the set of final states correspond to all available channels, i.e $\sigma$ is the total integral cross section, the CCS matrix elements can also be written as $ \mathcal{C}_{ij}^{tot}=\frac{4\pi}{k} Im[f_{i\rightarrow j}(\theta=0)]$, as imposed by the coherent multichannel optical theorem \cite{Devolder2022}.
In addition, the branching ratio between two different sets of final states (elastic versus reactive scattering for example) can be expressed as a generalized 
Rayleigh quotients of the corresponding CCS matrices $\boldsymbol{\mathcal{C}_1}$ and $\boldsymbol{\mathcal{C}_2}$:
	\begin{equation}
		r=\frac{\boldsymbol{a}^\dagger \boldsymbol{\mathcal{C}_1}\boldsymbol{a}}{\boldsymbol{a}^\dagger \boldsymbol{\mathcal{C}_2}\boldsymbol{a}}.
		\label{eq:brratio}
	\end{equation} 
Therefore, for any  scattering observable, any set of final states and any number of states in the initial superposition, coherent control can be expressed using the CCS matrix. This makes it a general theoretical framework for the study of coherent control. It is important to note that the CCS matrix contains explicit information about the property measurement in a scattering experiment. As such, the discussion of the experimental interference pattern is disctict from the traditional discussion of the density matrix \cite{Schlosshauer_book}. 

Given this generality, the following general notation is used in the next sections unless otherwise stated. We consider the contol of a scattering observable $O$, which is the sum of squared transition amplitudes $\left|o_{i\rightarrow f}\right|$ over $N_f$ final states (internal states, scattering angles, partial waves, etc ...):
\begin{equation}
	O=\sum_{f=1}^{N_f} \left|\sum_i^{N_i}a_i o_{i\rightarrow f} \right|^2
\end{equation}
and can be written as a Rayleigh quotient:
\begin{equation}
	O=\boldsymbol{a}^\dagger \boldsymbol{\mathcal{C}}\boldsymbol{a},
	\label{eq:cohe_ctrl_O}
\end{equation}
where the CCS matrix is defined as:
	\begin{equation}
	\mathcal{C}_{ij}=\sum_{f=1}^{N_f}o_{i\rightarrow f}o_{j\rightarrow f}^*.
	\label{eq:CCSM_gen}
\end{equation}
For example, for the differential cross section, the final state $\ket{f}$ corresponds to the final internal state $\ket{j'}$ plus the angles $(\theta,\phi)$, while for the integral cross section, the final state $\ket{f}$ corresponds to the final internal state $\ket{j'}$ plus the final partial wave $\ell'$.

\section{Analysis of the controllability with the CCS matrix}
The CCS matrix is a $N_i\times N_i$ Hermitian, positive semidefinite matrix, whose diagonal elements $\mathcal{C}_{ii}$ give the value of the scattering observable $O$ when the initial state is $\ket{i}$, and the off-diagonal element $\mathcal{C}_{ij}$ gives the interference contribution to $O$ induced by superposing the states $\ket{i}$ and $\ket{j}$. The CCS matrix takes into account the symmetries, which often results in its block-diagonalization. For example, when aiming to control the half-integral and integral cross sections, the CCS matrix is divided into blocks based on the projection of the total internal angular momentum, where there is no interference between states with different total internal projections \cite{Devolder2020,Omiste2018}.

\begin{figure}[t]
	\centering
	\includegraphics[width=\columnwidth]{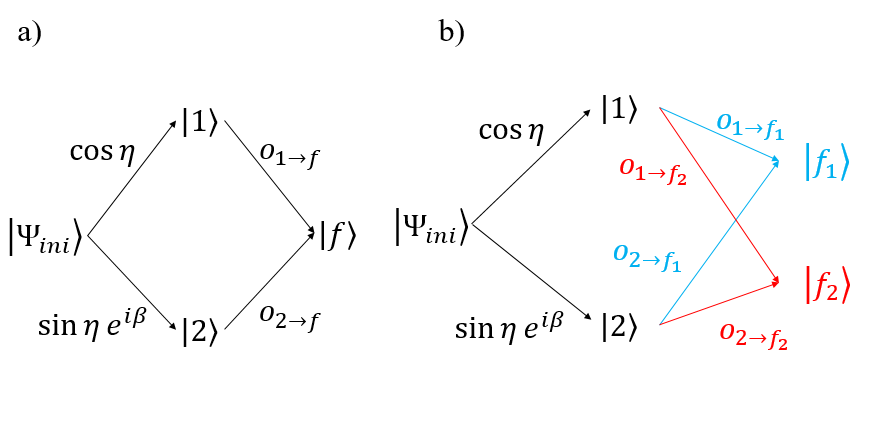}
	\caption{Analogy of coherent control with the double slit experiments for (a) 1 final state and (b) multiple final states. The parameters $\eta$ and $\beta$ are defined in Eq. (\ref{eq:sup_2states}).}
	\label{fig:Analogy}
\end{figure}

An analogy between coherent control of scattering and multiple slit experiments can be drawn. Note that the analogy has been developped in detail in \cite{Scholak2014,Scholak2017}. In this analogy, the basis states $\left\{\ket{i}\right\}$ play the role of the slits. There are two steps in the paths (see Fig. \ref{fig:Analogy}): first from the initial state to the basis states and secondly from the basis states to the final states.  The transition amplitudes from the initial state $\ket{\Psi_{ini}}$ to these basis states are given by the superposition coefficients $\begin{pmatrix} 
	a_1 & a_2 & ... & a_N
\end{pmatrix}$. This first step in coherent control can be controlled by changing the relative populations and phases between the different components ($\ket{1}$,$\ket{2}$, etc ...) of $\Psi_{ini}$. This is different from multiple-slit experiments, where the evolution is generally free in this first step. The transition amplitudes for the second step are given by the corresponding transition amplitudes $o_{i\rightarrow f}$, whom are specific to the studied system and are contained in the CCS matrix, constraining the controllability.
\subsection{2 superposition states, 1 final state}
To analyze controllability, we first consider the simplest case with two superposition states ($\ket{1},\ket{2}$) and one final state $\ket{f}$. The 2-state initial superposition is given by:
\begin{equation}
	\ket{\Psi_{ini}}=\cos\eta \ket{1}+ \sin\eta e^{i\beta}\ket{2},
	\label{eq:sup_2states}
\end{equation}
where $\eta=[0,\pi/2]$ and $\beta=[0,2\pi]$ determine, respectively, the relative population and phase. Two paths are created, one passing by $\ket{1}$  with the transition amplitude to $\ket{f}$ of $\cos\eta \ o_{1\rightarrow f}$ and one passing by $\ket{2}$ with the transition amplitude to $\ket{f}$ of $\sin\eta \ e^{i\beta} \ o_{2\rightarrow f}$ (see Fig. \ref{fig:Analogy} (a)). These two paths interfere and give an interference term in the scattering observable $O$:
\begin{equation}
	\begin{split}
		O=&\cos^2(\eta) |o_{1\rightarrow f}(\theta,\phi)|^2+ \sin^2(\eta) |o_{2\rightarrow f}(\theta,\phi)|^2\\&+ \cos\eta\sin\eta \ o_{1\rightarrow f}(\theta,\phi)o_{2\rightarrow f}^*(\theta,\phi)e^{-i\beta}\\ &+\cos\eta\sin\eta \ o_{1\rightarrow f}^*(\theta,\phi)o_{2\rightarrow f}(\theta,\phi)e^{i\beta}).
	\end{split}
\label{eq:cross_2state_O}
\end{equation}
As illustrated previously, this equation can be written instead in terms of the CCS matrix:
\begin{equation}
O=\cos^2(\eta) \mathcal{C}_{11}+	\sin^2(\eta)\mathcal{C}_{22}+ 2 \cos\eta\sin\eta |\mathcal{C}_{12}| \cos(\beta-arg(\mathcal{C}_{12}))
\label{eq:cross_2state},
\end{equation}
where $arg(\mathcal{C}_{12})$ is the argument of the off-diagonal element of the CCS matrix.

In coherent control, we wish to control a scattering observable $O$ by tuning the relative phase $\beta$ between the two superposition states. This is quantified by the visibility:
\begin{equation}
	V(\eta)=\frac{O_{max}^\eta-O_{min}^\eta}{O_{max}^\eta+O_{min}^\eta},
\end{equation}
where $O_{max}^\eta$ ($O_{min}^\eta$) is the maximal (minimal) value of the scattering observable $O$ when the relative population (given by $\eta$) are fixed and only the relative phase is varied. The maximization of visibility gives the best control. Using Eq. (\ref{eq:cross_2state}), the visibility is given by:
\begin{equation}
	\begin{split}
		V(\eta)=&\frac{2 \cos\eta\sin\eta \ |o_{1\rightarrow f}(\theta,\phi)|\ |o_{2\rightarrow f}(\theta,\phi)|}{\cos^2(\eta) |o_{1\rightarrow f}(\theta,\phi)|^2 +	\sin^2(\eta)|o_{2\rightarrow f}(\theta,\phi)|^2}\\=&\frac{2 \cos\eta\sin\eta |\mathcal{C}_{12}| }{\cos^2(\eta) \mathcal{C}_{11}+	\sin^2(\eta)\mathcal{C}_{22}}.	
	\end{split}
\label{eq:V}
\end{equation}
The visibility depends on the relative population parameter $\eta$ in Eq. (\ref{eq:sup_2states}). This $\eta$-dependence in the denominator indicates that the control tries to counterbalance the asymmetry between the transition amplitudes $|o_{1\rightarrow f}(\theta,\phi)|$ and $|o_{2\rightarrow f}(\theta,\phi)|$ and aims to decrease the distinguishability between the two paths. If $|o_{1\rightarrow f}(\theta,\phi)|$ is greater than $|o_{2\rightarrow f}(\theta,\phi)|$, a higher visibility is obtained for superposition with more relative population in the state $\ket{2}$, and vice versa. The visibility reaches its highest value for:
 \begin{equation}
 	\begin{split}		
	\eta_{V_{max}}=&\arctan\left(\frac{|o_{1\rightarrow f}(\theta,\phi)|}{|o_{2\rightarrow f}(\theta,\phi)|}\right)\\=&\arctan\left(\frac{\sqrt{\mathcal{C}_{11}}}{\sqrt{\mathcal{C}_{22}}}\right).
	 \end{split}
\label{eq:eta_maxV}
\end{equation}
By inserting eq. (\ref{eq:eta_maxV}) into Eq. (\ref{eq:V}), this highest value is found to be 1, with $O_{min}^\eta=0$. Therefore, in the case of a single final state, complete control and complete destructive interference are always achievable. As we will see later, this statement does not hold when more than one final states must be considered.

As for \textit{complementary}, we first define the path distinguishability from the initial state $\ket{\Psi_{ini}}$ to the final state. The path distinguishability is related to the asymmetry of the scattering observable $O$ obtained through the two paths. In a fictional scenario without interference, the first path would give $\cos^2\eta \ |o_{1\rightarrow f}(\theta,\phi)|^2$ while the second path  would give $\sin^2\eta \ |o_{2\rightarrow f}(\theta,\phi)|^2$ [see Eq. (\ref{eq:cross_2state_O})]. Therefore, the path distinguishability is defined as:
\begin{equation}
	\begin{split}
	P(\eta)=&\frac{\left|\cos^2\eta \ |o_{1\rightarrow f}(\theta,\phi)|^2-\sin^2\eta \ |o_{2\rightarrow f}(\theta,\phi)|^2\right|}{\cos^2\eta \ |o_{1\rightarrow f}(\theta,\phi)|^2+\sin^2\eta \ |o_{2\rightarrow f}(\theta,\phi)|^2}\\=&\frac{\left|\cos^2\eta \ 	\mathcal{C}_{11}-\sin^2\eta \ 	\mathcal{C}_{22}\right|}{\cos^2\eta \ \mathcal{C}_{11}+\sin^2\eta \ \mathcal{C}_{22}}.		
	\end{split}
	\label{eq:P}
\end{equation}
Note that the expression (Eq. \ref{eq:P}) can be rewritten in terms of the ratio between the arithmetic mean $\frac{\cos^2\eta \ \mathcal{C}_{11}+\sin^2\eta \ \mathcal{C}_{22}}{2}$ and the geometric mean $\sqrt{\cos^2\eta \ \mathcal{C}_{11} \sin^2\eta \ \mathcal{C}_{22}}$: 
\begin{equation}
		P(\eta)=\sqrt{1-\frac{4 \mathcal{C}_{11}\mathcal{C}_{22} \cos^2\eta \sin^2\eta}{\left(\cos^2\eta \ \mathcal{C}_{11} + \sin^2\eta \ \mathcal{C}_{22} \right)^2}}.
	\label{eq:P_mean}
\end{equation}
The arithmetic mean and the geometric means are only equal when $\cos^2\eta \ \mathcal{C}_{11}=\sin^2\eta \ \mathcal{C}_{22}$, resulting in zero path distinguishability. On the other hand, if either $\cos^2\eta \ \mathcal{C}_{11}$ or $\cos^2\eta \ \mathcal{C}_{11}$ is zero, the path distinguishability is equal to 1. This formulation (Eq. (\ref{eq:P_mean}))  will be important for the generalization to more than 2 initial superposition states. 

Like the visibility, the path distinguishability depends on the relative population. A special case is when $\eta=\pi/4$, where no bias is introduced in the initial superposition ($\frac{1}{\sqrt{2}}\left(\ket{1}+e^{i\beta}\ket{2}\right)$) and the path distinguishability gives the asymmetry between the transition amplitudes ($|o_{1\rightarrow f}(\theta,\phi)|$ and $|o_{2\rightarrow f}(\theta,\phi)|$):
\begin{equation}
\begin{split}
	P(\eta=\pi/4)=&\frac{\left||o_{1\rightarrow f}(\theta,\phi)|^2-|o_{2\rightarrow f}(\theta,\phi)|^2\right|}{|o_{1\rightarrow f}(\theta,\phi)|^2+|o_{2\rightarrow f}(\theta,\phi)|^2}\\=&\frac{\left|\mathcal{C}_{11}-\mathcal{C}_{22}\right|}{\mathcal{C}_{11}+\mathcal{C}_{22}}.
\end{split}
\end{equation}
In that case, the visibility is given by the interference $|\mathcal{C}_{12}|$  divided by the average of the diagonal elements $\mathcal{C}_{11}$ and $\mathcal{C}_{22}$:
\begin{equation}
\begin{split}
V (\eta=\pi/4)=&\frac{2|f_{1\rightarrow f}(\theta,\phi)|\ |f_{2\rightarrow f}(\theta,\phi)|}{|f_{1\rightarrow f}(\theta,\phi)|^2+|f_{2\rightarrow f}(\theta,\phi)|^2}\\=&\frac{2 |\mathcal{C}_{12}|}{\mathcal{C}_{11}+\mathcal{C}_{22}}.
\end{split}
\end{equation}

A complementary relation holds between the visibility and the path distinguishability:
\begin{equation}
\begin{split}
P^2(\eta)+V^2(\eta)=&\frac{(\cos^2\eta \ \mathcal{C}_{11}-\sin^2\eta \ \mathcal{C}_{22})^2}{(\cos^2\eta \ \mathcal{C}_{11}+\sin^2\eta \ \mathcal{C}_{22})^2}\\+&\frac{4\cos^2\eta\sin^2\eta |\mathcal{C}_{12}|^2 }{(\cos^2\eta \ \mathcal{C}_{11}+\sin^2\eta \ \mathcal{C}_{22})^2}.
\end{split}
\label{eq: compl_rel_1}
\end{equation}
As $|\mathcal{C}_{12}|=\sqrt{\mathcal{C}_{11}\mathcal{C}_{22}}$ for one final state, the usual relation between visibility and path distinguishability is recovered:
\begin{equation}
\begin{split}
	P^2(\eta)+V^2(\eta)=&\frac{(\cos^2\eta \ \mathcal{C}_{11}-\sin^2\eta \ \mathcal{C}_{22})^2}{(\cos^2\eta \ \mathcal{C}_{11}+\sin^2\eta \ \mathcal{C}_{22})^2}\\+&\frac{4\cos^2\eta\sin^2\eta \mathcal{C}_{11}\mathcal{C}_{22}  }{(\cos^2\eta \ \mathcal{C}_{11}+\sin^2\eta \ \mathcal{C}_{22})^2}\\=&1.	
\end{split}
	\label{eq: compl_rel_2}
\end{equation}
Equation (\ref{eq: compl_rel_2}) indicates that a decrease in distinguishability induces an increase in visibility and vice versa. First, this can be achieved through a change in the relative population in the superposition. For $\eta=\eta_{V_{max}}$, the visibility is equal to one, while the distinguishability is zero. On the other hand, for $\eta=0$ and $\pi/2$, the initial states occupies only one state. Therefore, the distinguishability is 1 and the visibility is zero. Between these two extremes, the visibility and the path distinguishability have values between 0 and 1, respecting the complementary relation in Eq.(\ref{eq: compl_rel_2}). Secondly, the visibility and the distinguishability depend on the elements of the CCS matrix. An example of this is a change of basis, explored in this paper. 
\subsection{2 superposition states, multiple final states}
The beauty of the CCS formalism is that the expressions for the cross section (Eq.(\ref{eq:cross_2state})), the visibility (Eq. \ref{eq:V}) and the path distinguishability (Eq.(\ref{eq:P})) in terms of the CCS matrix elements remain the same for multiple final states. An additional advantage of the CCS formalism is computational. Indeed, it is easier to first calculate the CCS matrix and then obtain all the quantities of interest.

The major difference between the single final states and the multiple final state cases is that the relation $|\mathcal{C}_{12}|=\sqrt{\mathcal{C}_{11}\mathcal{C}_{22}}$ no longer holds. Instead, it is replaced by the Schwartz inequality, $|\mathcal{C}_{12}|\le\sqrt{\mathcal{C}_{11}\mathcal{C}_{22}}$. This difference implies that the maximal visibility is no longer 1 and is limited by this Schwartz inequality: 
\begin{equation}
	V(\eta)= R_c\frac{2 \cos\eta\sin\eta \sqrt{\mathcal{C}_{11}\mathcal{C}_{22}}}{\cos^2(\eta) \mathcal{C}_{11}+	\sin^2(\eta)\mathcal{C}_{22}},
	\label{eq:V_multiple}
\end{equation}
where we have introduced the control index
\begin{equation}
R_c=\frac{|\mathcal{C}_{12}|}{\sqrt{\mathcal{C}_{11}\mathcal{C}_{22}}}
\end{equation}
 which corresponds to the maximal obtainable visibility. The control index $R_c$ quantifies the controllability of the system and is related to the difference of control landscapes between the final states. When the controls of every final state are synchronized \cite{Devolder2023}, $R_c$ is close to 1 and complete destructive interference is obtained. On the other hand, when the individual controls are asynchronous, the global control is hindered by the competition between the individual controls and the control index $R_c$ has a low value compared to 1. 

Another consequence of the Schwartz inequality is in the complementary relation between the visibility and the path distinguishability. The transition from the equ.(\ref{eq: compl_rel_1}) to equ.(\ref{eq: compl_rel_2}) is no longer possible. The complementary relation is constrained by the control index: 
\begin{widetext}
\begin{equation}
	P^2(\eta)+V^2(\eta)=\frac{(\cos^2\eta \ \mathcal{C}_{11}-\sin^2\eta \ \mathcal{C}_{22})^2+4\cos^2\eta\sin^2\eta |\mathcal{C}_{12}|^2 }{(\cos^2\eta \ \mathcal{C}_{11}+\sin^2\eta \ \mathcal{C}_{22})^2}
	\label{eq: compl_rel_3}
\end{equation}
\begin{equation}
	P^2(\eta)+V^2(\eta)=\frac{(\cos^2\eta \ \mathcal{C}_{11}-\sin^2\eta \ \mathcal{C}_{22})^2+4\cos^2\eta\sin^2\eta R_c^2 \mathcal{C}_{11}\mathcal{C}_{22} }{(\cos^2\eta \ \mathcal{C}_{11}+\sin^2\eta \ \mathcal{C}_{22})^2}
	\label{eq: compl_rel_4}
\end{equation}
\begin{equation}
	P^2(\eta)+V^2(\eta)=1+\frac{4\cos^2\eta\sin^2\eta \mathcal{C}_{11}\mathcal{C}_{22} (R_c^2-1) }{(\cos^2\eta \ \mathcal{C}_{11}+\sin^2\eta \ \mathcal{C}_{22})^2} 
	\label{eq: compl_rel_5}
\end{equation}
\end{widetext}
Since $R_c$ has a value between 0 and 1, the second term is always negative and therefore, the sum of squared quantities is bounded by 1:
\begin{equation}
	P^2(\eta)+V^2(\eta)=1+\frac{4\cos^2\eta\sin^2\eta \mathcal{C}_{11}\mathcal{C}_{22} (R_c^2-1) }{(\cos^2\eta \ \mathcal{C}_{11}+\sin^2\eta \ \mathcal{C}_{22})^2} \le 1.
	\label{eq: compl_rel_6}
\end{equation}
A good example of this bounded complementary relation is for $	\eta_{V_{max}}=\arctan\left(\frac{\sqrt{\mathcal{C}_{11}}}{\sqrt{\mathcal{C}_{22}}}\right)$ where the visibility is maximal and is equal to $R_c$. However, the path distinguishability is still zero, as with a single final state. Therefore, $P^2(\eta_{V_{max}})+V^2(\eta_{V_{max}})$ is equal to $R_c^2$, which is smaller or equal to 1. 
\subsection{General case: multiple superposition states, multiple final states}
The formulation based on the CCS matrix is particularly suitable to take one step further and  generalize to an arbitrary number of superposition states and an arbitrary number of final states. First, the general form for the scattering observable $O$ is given by the Eq. (\ref{eq:cohe_ctrl_O}). The visibility is directly related to the extent of control only when the $N_i-1$ relative phases are tuned and the relative populations are fixed. Therefore, as in the case of 2 superposition states considered above,  the visibility is a function of all relative population parameters $|a_i|$ in the general case and is given by the ratio between the contributions due to the interference and incoherent terms:
\begin{equation}
	V(\{|a_i|\})=\frac{\sum_{i,j}|a_i||a_j| |\mathcal{C}_{ij}|}{\sum_i |a_i|^2 \mathcal{C}_{ii}}.
\end{equation}
For the path distinguishability, we generalize the expression (Eq. \ref{eq:P_mean}) with the geometric and arithmetic means of $|a_i|^2 \mathcal{C}_{ii}$:
\begin{equation}
		P(\{|a_i|\})=\sqrt[N]{1-\frac{N^N \prod_i |a_i|^2 \mathcal{C}_{ii} }{\left(\sum_i |a_i|^2 \mathcal{C}_{ii}\right)^N}}.
\end{equation}
With this definition, $P(\{|a_i|\})$ is zero when all $|a_i|^2 \mathcal{C}_{ii}$ are equal, as expected.

Finally, a general formula can be given for the control index $R_c$, which quantifies the controllability. The control index is equal to 1 when the minimum value of the cross section is zero, i.e the case of complete destructive interference. Since the equation for the scattering observable $O$ (Eq. (\ref{eq:cohe_ctrl_O})) is written as a Rayleigh quotient, the last statement means that the lowest eigenvalue of the CCS matrix must be equal to zero for complete control \cite{Frishman1999,Brumer_book}.
This can be determined by checking if the determinant of the CCS matrix is zero. Frishmann et al. \cite{Frishman1999} demonstrated that this is achieved when $N_i>N_f$. On the other hand, the lack of interference and hence of interference-based
control ($R_c=0$) is characterized by a diagonal CCS matrix, whose the determinant is equal to the product of the diagonal elements. Therefore, the control index $R_c$ can be defined based on the determinant of the CCS matrix:
\begin{equation}
	R_c=\sqrt[N]{1-\frac{det(\boldsymbol{\mathcal{C}})}{\prod_i \mathcal{C}_{ii}}}.
	\label{eq:cont_index_N}
\end{equation}
$R_{c}$ lies between 0 ($det(\boldsymbol{\mathcal{C}})=\prod_i \mathcal{C}_{ii}$) and 1 ($det(\boldsymbol{\mathcal{C}})=0$). For a 2-state initial superposition, $det(\boldsymbol{\mathcal{C}})=\mathcal{C}_{11} \mathcal{C}_{22}-|\mathcal{C}_{12}|^2$  and we recover  $R_c=\frac{|\mathcal{C}_{12}|}{\sqrt{\mathcal{C}_{11} \mathcal{C}_{22}}}$.

\section{Exposing interference through a change of basis}

The CCS matrix formalism of coherent control is well-suited for examining the impact of a change of basis on quantum interference. Here, we present the general theory of basis 
transformations, with our initial focus on orthogonal transformations. In particular, we demonstrate the appearance or disappearance of interference contributions depending on the basis.
\subsection{Orthogonal transformation}
Consider two sets of basis states $\{\ket{i}\}$ and $\{\ket{j'}\}$ related by the unitary transformation $\boldsymbol{U}$:
	\begin{equation}
		\ket{i}=\sum_{j'}U_{j'i} \ket{j'}.
		\label{eq;ini_exp}
	\end{equation}
The initial states $\ket{\Psi_{ini}}$ can be expressed as a superposition in both bases:
	\begin{equation}
		\ket{\Psi_{ini}}=\sum_i a_i \ket{i}=\sum_{j'}d_{j'} \ket{j'},
	\end{equation}
and the coefficients are related by $\boldsymbol{U}$:
	\begin{equation}
		\boldsymbol{d}=\boldsymbol{U}\boldsymbol{a}.
	\end{equation}
The scattering observable $O$ (Eq.(\ref{eq:cohe_ctrl_O})) can be expressed in the new basis as:
	\begin{equation}
		O=\boldsymbol{d}^\dagger \boldsymbol{\mathcal{C}'}\boldsymbol{d},
		\label{eq:diff_basis}
	\end{equation}
	where the CCS matrices in the new basis are simply related to the CCS matrices in the old basis by:
	\begin{equation}
		\boldsymbol{\mathcal{C}'}=\boldsymbol{U}\boldsymbol{\mathcal{C}}\boldsymbol{U}^\dagger.
		\label{eq:F_basis}
	\end{equation}
	
Modifying the interference and visibility through a change of basis is related to the non-commutativity of the transformation matrix $\boldsymbol{U}$ and the CCS matrix $\boldsymbol{\mathcal{C}}$. When the two matrices commute, the CCS matrix remains unchanged in the new basis and the visibility is identical. Otherwise, if they do not commute, the visibility would change. Taking the example of the 2-state superposition, this change can manifest in two ways. First, the asymmetry of the transition amplitudes $|o_{1'\rightarrow f}(\theta,\phi)|$ and $|o_{2'\rightarrow f}(\theta,\phi)|$ can be modified via a change of basis, inducing a change of the $\eta$-dependance of the visibility $V(\eta)$ (Eq. \ref{eq:V}) and of the path distinguishability $P(\eta)$ (Eq. \ref{eq:P}). As a reminder, the $\eta$-dependence of the visibility $V(\eta)$ illustrates the counterbalance strategy of the control against the asymmetry of the transition amplitudes. If $\mathcal{C}_{11}>\mathcal{C}_{22}$ becomes $\mathcal{C}'_{11}<\mathcal{C}'_{22}$ in the new basis, the higher values of the visibility would shift from the interval $[\pi/4,\pi/2]$ to the interval $[0,\pi/4]$, and vice versa. Secondly, if the scattering observable $O$ is the sum over multiple final states, the maximal value of the visibility, i.e the control index $R_c$, would be modified. 

An illustrative example is when there is no interference and no coherent control ($R_c=0$) in the initial basis, i.e the CCS matrix is diagonal in that basis. An interference structure will appear in the transformed basis if the transformation $\boldsymbol{U}$ does not commute with the diagonal CCS matrix. This condition is fullfilled if the new basis states are superposition of initial basis states giving different values of the scattering observable $O$. Therefore, if the diagonal elements of the initial CCS matrix are all different, any non-diagonal transformation $\boldsymbol{U}$ can induce interference and increase the control index $R_c$. On the other hand, if all diagonal elements of the CCS initial matrix are identical, i.e the CCS matrix is proportional to the identity matrix, a change of basis cannot lead to the appearance of interference and the control index remains zero.

\subsection{Suppression of interference: The eigenvector basis}	
Diagonalization of the CCS matrix is a special orthogonal transformation $\boldsymbol{U_{diag}}$ insofar as it eliminates the interference structure in the new basis.
This transformation is closely connected to the optimization of the scattering observable $O$ \cite{Frishman1999}, since the Eq. (\ref{eq:cohe_ctrl_O}) is expressed as a Rayleigh quotient. Indeed, the minimal $O_{min}$ and maximal $O_{max}$ values of the scattering observable $O$ are determined by the  smallest and largest eigenvalues of the CCS matrix. (Note that the positive semidefinite character of the CCS matrix guarantees that the minimal value of the scattering observable $O$ is non-negative). The maximum and minimum values of the scattering observable $O$ are the same in any basis. The difference between the basis is in the decomposition of the corresponding eigenvectors $\ket{max}$ and $\ket{min}$. Due to the hermiticity of the CCS matrix, these eigenvectors must be orthogonal, which imposes some constraints. For example, for a 2-state initial superposition ($\ket{\Psi_{ini}}=\cos\eta \ket{1}+ \sin\eta e^{i\beta}\ket{2}$), the superposition parameters for the minimum and maximal values are related by $\eta_{max}+\eta_{min}=\pi/2$ and $\beta_{max}-\beta_{min}=\pi$.

The optimization through diagonalization of the CCS matrix explores the entire Hilbert space spanned by the $N_i$ initial states. If the internal structure of both colliding particles is considered in the superposition, diagonalization typically results in entangled superpositions \cite{Devolder2021} as extreme eigenvectors. While it is logical that entanglement can enhance control, it is unlikely that the two colliding particles have already interacted before the collision and are entangled. Therefore, in most cases, the optimization should focus on exploring the subspace of non-entangled superpositions, and the range of control corresponds to the so-called numerical product range of the CCS matrix \cite{Gawron2010}. 

Due to the hermiticity of the CCS matrix, one can always define a basis of its eigenvectors in which the scattering observable $O$ (Eq. (\ref{eq:cross_2state})) will be free of interference contributions. In such basis, control of $O$ can be achieved by variations in the relative populations since there are no terms depending on relative phases. The path distinguishability $\mathcal{P}$ is maximized. Even though the eigenvector basis is always defined, it may not be the most appropriate basis for several reasons. First, it can be difficult to realise experimentally. Secondly, different final states corresponds to different CCS matrices and, consequently, different eigenvectors, except for some cases constrained by symmetry. For example, control of the differential cross section at one angle ($\theta,\phi$) may be interference-free in an eigenvector basis, but interferences may  contribute at other angles. Similarly, for example,  a collisional process that includes elastic and inelastic scattering may be interference-free for inelastic
scattering in  an basis  but interference-sensitive for elastic scattering in that same basis. 

This circumstance is particularly significant for control of a branching ratio $r$. In general, a unitary transformation $\boldsymbol{U}$ can change the branching ratio as the generalized Rayleigh quotient with the CCS matrices $\boldsymbol{\mathcal{C}_1'}$ and $\boldsymbol{\mathcal{C}_2'}$ in the new basis, given by
\begin{equation}
	r=\frac{\boldsymbol{d}^\dagger \boldsymbol{\mathcal{C}_1'}\boldsymbol{d}}{\boldsymbol{d}^\dagger \boldsymbol{\mathcal{C}_2'}\boldsymbol{d}}.
\end{equation}  
If the CCS matrices $\boldsymbol{\mathcal{C}_1}$ and $\boldsymbol{\mathcal{C}_2}$ do not commute, the eigenvectors for the two CCS matrices will differ, leading to interference for at least one of the set of final states, regardless of the basis chosen. Therefore,\textit{controlling the branching ratio can not in general be turned into interference-free control by a change of basis.}

While the optimization of a Rayleigh quotient corresponds to solve an eigenvalue problem, the determination of the minimal and maximal value for the branching ratio as a generalized Rayleigh quotient is given by the generalized eigenvalue problem
$\boldsymbol{\mathcal{C}_1} \boldsymbol{a}=\lambda \boldsymbol{\mathcal{C}_2} \boldsymbol{a}$ 
(from Eq. (\ref{eq:brratio}) with the two corresponding CCS matrices \cite{Zhang_book}. As previously, the minimal and maximal values of the branching ratio remain invariant to any change of basis.

\subsection{Maximizing interference: The Fourier basis}

While diagonalization minimizes the control index $R_c$ and maximizes the path distinguishability, there is also an alternate transformation that maximizes the control index $R_c$ and minimizes the path distinguishability. In the same way that $R_c$ is zero in the eigenbasis, the path distinguishability would be zero in the maximal interference basis. One achieves this by making all the diagonal elements of the CCS matrix equal in this basis. This transformation corresponds to applying the quantum Fourier transform \cite{Yao2016,Hu2017} to the eigenbasis. Therefore, the complete transformation from any basis to the Fourier basis requires diagonalization of the CCS matrix in the initial basis followed by the quantum Fourier transform. The matrix elements of the quantum Fourier transform are given by:
\begin{equation}
	[QFT]_{jk}=\frac{1}{\sqrt{N_i}}\exp\left(\frac{ 2\pi i (j-1)(k-1) }{N_i}\right).
	\label{eq:QFT_transf}
\end{equation} 

For a superposition of two states, the quantum Fourier transform reduces to the Hadamard transformation:
\begin{equation}
	QFT=H=\frac{1}{\sqrt{2}}\begin{pmatrix}
		1 & 1 \\ 1 & -1	
	\end{pmatrix}.
\end{equation}
The states of the maximal interference basis are the equal superpositions of the eigenvectors:
\begin{equation}
	\ket{int_1}=\frac{1}{\sqrt{2}}\left(\ket{max}+\ket{min}\right)
\end{equation}
\begin{equation}
	\ket{int_2}=\frac{1}{\sqrt{2}}\left(\ket{max}-\ket{min}\right).
\end{equation}
The application of the Hadamard Gate to a diagonal CCS matrix results in:
\begin{equation}
	\mathcal{C}'=
	\frac{1}{2}\begin{pmatrix}
		1 & 1 \\ 1 & -1	
	\end{pmatrix}
	\begin{pmatrix}
		O_{min} & 0 \\ 0 & O_{max}	
	\end{pmatrix}
	\begin{pmatrix}
		1 & 1 \\ 1 & -1	
	\end{pmatrix}
\end{equation}

\begin{equation}
	\mathcal{C}'=\frac{1}{2}\begin{pmatrix}
		O_{max}+O_{min} & O_{min}-O_{max} \\ O_{min}-O_{max} & O_{max}+O_{min}	
	\end{pmatrix}.
\end{equation}
As expected, the diagonal elements of the CCS matrix are equal. The control index $R_c$ in the Fourier basis is given by :
\begin{equation}
	R_c=\frac{O_{max}-O_{min}}{O_{max}+O_{min}},
	\label{eq:int_max_basis}
\end{equation}
which is the maximal achievable visibility we can obtain through a change of basis. This maximal visibility in the Fourier basis is obtained for $\eta=\pi/4$, i.e the equal population superpositions $\frac{1}{\sqrt{2}}\left(\ket{int_1}+e^{i\beta}\ket{int_2}\right)$ for which the path distinguishability is zero. The control index can be tuned from 0 in the eigenvector basis to $\frac{O_{max}-O_{min}}{O_{max}+O_{min}}$ in the Fourier basis. Its value in any other orthogonal basis would be between these two extremes.

\subsection{Non-orthogonal transformation}

This subsection extends the treatment to transformations involving non-orthogonal bases, such as  the
 $\{\ket{\psi_+}, \ket{\psi_-}\}$ basis used in the He + D$_2$ experiment \cite{Zhou2021}
discussed later below.  The fact that the non-orthogonal basis vectors have non-zero overlap has profound consequences for the issue of path indistinguishability and hence interferences.

Consider the transformation from an orthogonal basis $\ket{i}$ to a non-orthogonal basis $\ket{j'}$, characterized by the transformation matrix $\boldsymbol{P}$:
	\begin{equation}
		\ket{i}=\sum_{j'}P_{j'i} \ket{j'}.
	\end{equation}
	
A general non-unitary transformation can be expressed as the product of a unitary matrix
 $\boldsymbol{U} $
 and the square root of the overlap matrix, denoted as $S^{ovlp}_{i'j'}=\braket{i'|j'}$ (which is a Hermitian matrix):
	\begin{equation}
	\boldsymbol{P}=(\boldsymbol{S^{ovlp}})^{-1/2} \boldsymbol{U}.
	\end{equation}
Consequently, the product of the transformation matrix $\boldsymbol{P}$ with its transpose conjugate is no longer the identity matrix but rather is the inverse of the overlap matrix:
	\begin{equation}
		\boldsymbol{P} \boldsymbol{P}^\dagger=(\boldsymbol{S^{ovlp}})^{-1}.
		\label{eq. identity rel}
	\end{equation}
	
The initial state $\ket{\Psi_{ini}}$ can be expressed in either basis as in Eq. (\ref{eq;ini_exp}) with the following relations between the coefficients:
	\begin{equation}
		\boldsymbol{d}=\boldsymbol{P}\boldsymbol{a}
	\end{equation}
	or 
	\begin{equation}
		\boldsymbol{a}=\boldsymbol{P}^\dagger\boldsymbol{S^{ovlp}}\boldsymbol{d}=\boldsymbol{U}^\dagger(\boldsymbol{S^{ovlp}})^{1/2}\boldsymbol{d}.
		\label{eq;basis_coeff}
	\end{equation} 
To derive the matrix formulation of interference and of coherent control in a non-orthogonal basis we start from the Rayleigh quotient $\frac{\boldsymbol{a}^\dagger \boldsymbol{\mathcal{C}}\boldsymbol{a}}{\boldsymbol{a}^\dagger\boldsymbol{a}}$ and examine how the numerator and denominator change under the transformation. \\ For the numerator, we have:
	\begin{align}
		\boldsymbol{a}^\dagger \boldsymbol{\mathcal{C}}\boldsymbol{a}
		=& \boldsymbol{d}^\dagger \boldsymbol{S^{ovlp}} \boldsymbol{P}\boldsymbol{\mathcal{C}}\boldsymbol{P}^\dagger\boldsymbol{S^{ovlp}} \boldsymbol{d} \label{eq:transf_non-orth} \nonumber\\
		=&	\boldsymbol{d}^\dagger (\boldsymbol{S^{ovlp}})^{1/2}\boldsymbol{U}\boldsymbol{\mathcal{C}}\boldsymbol{U}^\dagger (\boldsymbol{S^{ovlp}})^{1/2}\boldsymbol{d} \nonumber \\
		=& \boldsymbol{d}^\dagger \boldsymbol{\mathcal{C}'}  \boldsymbol{d}.
	\end{align}
	Here, \begin{equation}
	\boldsymbol{\mathcal{C}'}=(\boldsymbol{S^{ovlp}})^{1/2}\boldsymbol{U}\boldsymbol{\mathcal{C}}\boldsymbol{U}^\dagger (\boldsymbol{S^{ovlp}})^{1/2}	
	\label{eq:CCS_mat_nonort}
	\end{equation} is the CCS matrix in the non-orthogonal basis. \\ For the denominator, Eq. (\ref{eq;basis_coeff}) yields:
	\begin{equation}
		\boldsymbol{a}^\dagger \boldsymbol{a}=\boldsymbol{d}^\dagger \boldsymbol{S}^{ovlp} \boldsymbol{d}.
	\end{equation}
Then, the matrix formulation of coherent control in a non-orthogonal basis becomes a generalized Rayleigh quotient:
\begin{equation}
O=\frac{\boldsymbol{d}^\dagger \boldsymbol{\mathcal{C}'} \boldsymbol{d}}{\boldsymbol{d}^\dagger \boldsymbol{S}^{ovlp} \boldsymbol{d}}.
\end{equation}
Even if $\boldsymbol{a}^\dagger \boldsymbol{a}=\boldsymbol{d}^\dagger \boldsymbol{S}^{ovlp} \boldsymbol{d}=1$, writing $\boldsymbol{d}^\dagger \boldsymbol{S}^{ovlp} \boldsymbol{d}$ in the denominator is useful to keep in mind this constraint. For example, the maximum and minimum values of the scattering observable $O$ are determined by solving the generalized eigenvalue problem $\boldsymbol{\mathcal{C}'} \boldsymbol{d}=\lambda \boldsymbol{S}^{ovlp} \boldsymbol{d}$ using the transformed CCS matrix and the overlap matrix, and not solely the eigenvalue problem with the transformed CCS matrix.

As discussed above, the impact of the non-orthogonal transformation on the interference and the controllability 
depends on the non-commutativity of  $\boldsymbol{P}$ 
with the CCS matrix. Mixing states with different values of the scattering observable $O$ will also change the interference structure in the new basis. Compared to the case of the orthogonal basis, the new feature for non-orthogonal basis sets is the significant effect of $(\boldsymbol{S^{ovlp}})^{-1/2}$. Even if $\boldsymbol{U}$ and the CCS matrix commute, the CCS matrix is still affected by the non-orthogonality of the new basis, since it follows from Eq. (\ref{eq:CCS_mat_nonort}) that  $\boldsymbol{\mathcal{C}'}=(\boldsymbol{S^{ovlp}})^{1/2}\boldsymbol{\mathcal{C}}(\boldsymbol{S^{ovlp}})^{1/2}$. Since the overlap matrix is non-diagonal by definition, the non-orthogonality of the new basis gives rise to additional interference. This encapsulates the complementarity relationship between amount of interference $\mathcal{I}$ and the path distinguishability $\mathcal{P}$. Specifically, non-orthogonality makes it more challenging to distinguish the states in the basis. Consequently, as path distinguishability decreases, the amount of interference increases.

\section{Controllability in different bases: example of $^{85}$Rb+ $^{85}$Rb scattering}
The theory of the coherent control in different bases developed above is illustrated here with a 2-state superposition case. The scattering between two $^{85}$Rb atoms in their hyperfine ground states $F=2$ is considered as an example. We have studied this case previously \cite{Devolder2022} and the details of the calculation of the S-matrix elements are given in that paper. We focus on the transitions to the final states $\ket{F=2,M_F=-2} \ket{2,2}$ and $\ket{2,0} \ket{2,0}$ of two Rb atoms. To be clear, the integral cross section considered in this example is the sum of the integral cross sections to the $\ket{2,-2} \ket{2,2}$ and $\ket{2,0} \ket{2,0}$ final states. 

\begin{figure}
	\centering
	\includegraphics[width=\columnwidth]{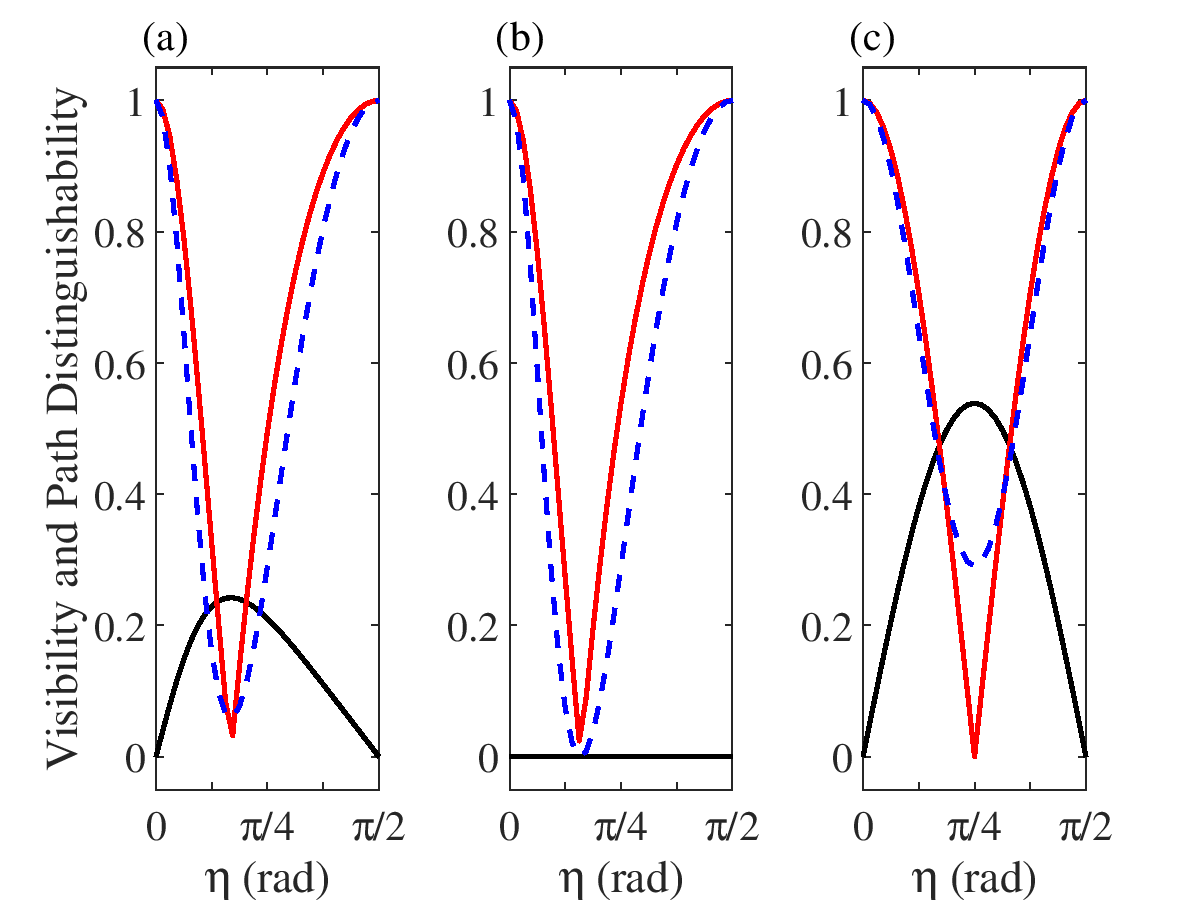}
	\caption{$\eta$-dependence of the visibility (black full line), the path distinguishability (red full line) and the sum $P^2(\eta)+V^2(\eta)$ (blue dashed line) for $^{85}$Rb+ $^{85}$Rb scattering in three different bases: (a) m-state basis, (b) CCS eigenvector basis and (c) Fourier basis}
	\label{fig:Basis_visibility}
\end{figure}

We consider the initial superposition of the states $\ket{2,-1} \ket{2,1}$ and $\ket{2,0}\ket{2,0}$, which constitute our first (orthogonal) basis, termed the m-state basis. This is a natural basis to consider, as it is composed of eigenstates of the two non-interacting atoms in the absence of external fields. The CCS matrix in this $m$-state basis is:
 \begin{equation}
 	\mathcal{C}=\begin{pmatrix}
 		11759 & 92-4909i \\
 		92+4909i & 34896
 	\end{pmatrix}
 \end{equation}
\begin{figure}
	\centering
	\includegraphics[width=\columnwidth]{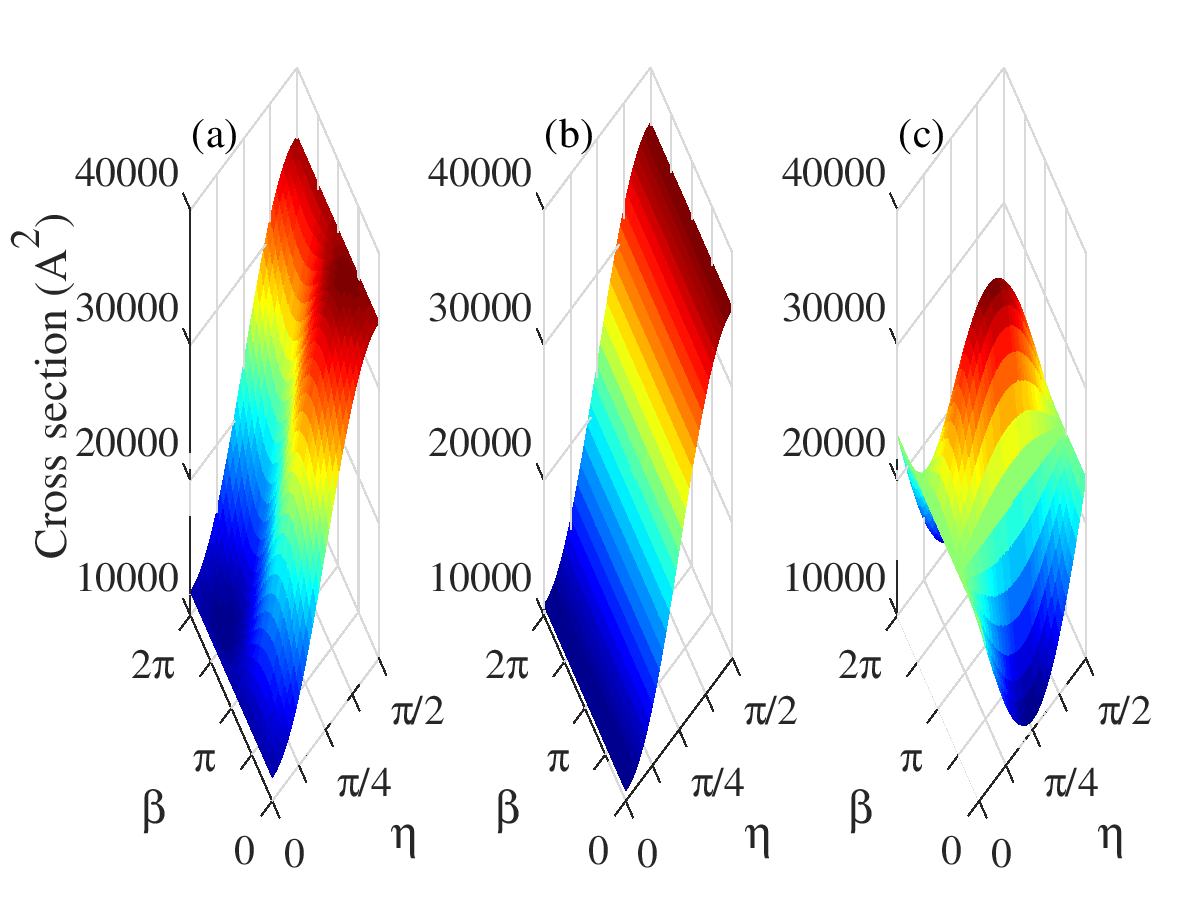}
	\caption{Coherent control landscape for $^{85}$Rb+ $^{85}$Rb scattering in  three different bases: (a) $m$-state basis, (b) CCS eigenvector basis and (c) Fourier basis. The cross section being controlled is to the $\ket{2,-2} \ket{2,2}$ and $\ket{2,0} \ket{2,0}$ final states and the collision energy is 50 $\mu$K.}
	\label{fig:Basis_landscape}
\end{figure}
The control landscape is depicted in Fig. \ref{fig:Basis_landscape} (a), illustrating the strong asymmetry between the cross sections from the two initial basis states (11759 $\AA^2$ from $\ket{2,-1} \ket{2,1}$ versus 34896 $\AA^2$ from $\ket{2,0}\ket{2,0}$). This asymmetry also affects the $\eta$-dependance of the visibility and the path distinguishability, as shown in Fig. \ref{fig:Basis_visibility} (a). Since the cross section is higher for the initial state $\ket{2,0}\ket{2,0}$, optimal control is achieved by increasing the population in the initial state  $\ket{2,-1} \ket{2,1}$ and decreasing the path distinguishability. 

The minimal and maximal values of the cross section are 10760 and 35895 $\AA^2$ which closely resemble the cross sections without superposition (11759 $\AA^2$ from $\ket{2,-1} \ket{2,1}$ and 34896 $\AA^2$ from $\ket{2,0}\ket{2,0}$). This similarity indicates a minimal  amount of interference and limited control extent. Indeed, the highest visibility, i.e, the control index, is only 0.24. Note that at $\eta_{V_{max}}$, the value of $\eta$ where the visibility is maximal, the path disitinguishability is minimal as expected and the sum $P^2(\eta_{V_{max}})+V^2(\eta_{V_{max}})$ is lower than 1.  

As a second basis consider the CCS eigenvector basis. The transformation to this basis is obtained by diagonalizing the CCS matrix:
 \begin{equation}
	\mathcal{C}_{diag}=\begin{pmatrix}
		10760 & 0 \\
		0 & 35895
	\end{pmatrix}.
\end{equation}
As noted above, the diagonal character of the CCS matrix  implies there is no intereference in the CCS eigenvector basis. The visibility is zero at any value of $\eta$, as shown in Fig. \ref{fig:Basis_visibility} (b) , illustrating that coherent control is impossible in this basis. The only possible control is classical as shown by the lack of the $\beta$-dependence in the control landscape [see Fig.\ref{fig:Basis_landscape} (b)]. Note that the maximal and the minimal values are still the same in this basis and corresponds to the cross sections obtained from the eigenvectors  $\ket{min}$ and $\ket{max}$.

Finally, we consider the Fourier basis as an extreme opposite of the CCS eigenvector basis. This basis is obtained by applying the Hadamard transformation to the eigenvector basis, giving the following CCS matrix:
 \begin{equation}
	\mathcal{C}_{Fourier}=\begin{pmatrix}
		23327 & -12568 \\
		-12568 & 23327
	\end{pmatrix}.
\end{equation}
This transformation resymmetrizes the control landscape (see Fig. \ref{fig:Basis_landscape} (c)). The cross-sections from $\ket{int_1}$ and $\ket{int_2}$ are identical. The highest visibility is achievable for $\eta=\pi/4$ (equal superpositions of the form $\frac{1}{\sqrt{2}}\left(\ket{int_1}+e^{i\beta}\ket{int_2}\right)$) and reaches 0.54. The control index doubles in going from the m-state basis to the Fourier basis. Furthermore, thanks to this symmetrization, the minimal (10760 $\AA^2$) and the maximal (35895 $\AA^2$) values of the cross section occur at $\eta=\pi/4$. Therefore, the minimal and the maximal values can be connected by tuning the relative phase $\beta$ without changing the relative population. This in contrast with the eigenvector basis, where the minimal and the maximal values are connected by varying the relative populations. 
\section{Analysis of He + D$_2$ Scattering}

A recent inelastic scattering experiment \cite{Zhou2021} on He + D$_2$  provides an enlightening example
of the revival of interference through a change of basis.  In these experiments, superpositions of degenerate magnetic sub-levels 
of D$_2 (v=2, j=2)$ rotational states ($m$-superpositions \cite{Brumer1999}) were prepared using Stark-induced adiabatic Raman passage (SARP) \cite{Mukherjee2014,Perreault2019a,Perreault2017,Perreault2018,Perreault2019b,Perreault2021,Zhou2021,Zhou2022}. These superpositions were subsequently employed in cold He + D$_2$ collision experiments, in which the inelastic differential cross section $\sigma(\theta)$ to the v=2 ,j'=0 states of D$_2$ were measured.

\begin{figure}
	\centering
	\includegraphics[width=\columnwidth]{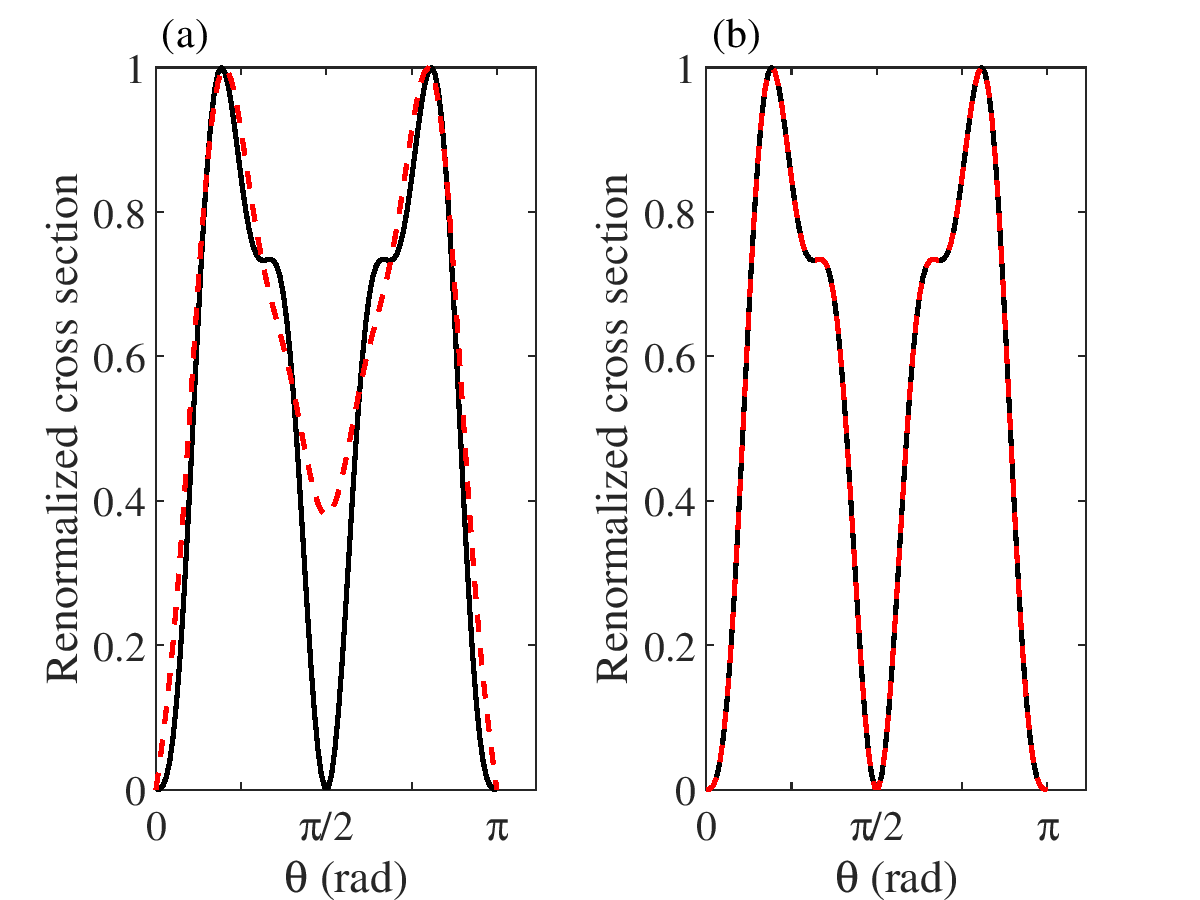}
	\caption{Cross section obtained from the S-matrix fit to experiment \cite{Zhou2021} (Normalized with respect to the highest value). Panel (a) shows the cross section from the $\ket{\psi_X}$ (black line) and the sum of the cross sections from $\ket{\psi_+}$ and $\ket{\psi_-}$ (red dashed line). Panel (b) shows the cross section from the $\ket{\psi_X}$ (black line) and the sum of the cross sections from $\ket{m=+1}$ and $\ket{m=-1}$ (red dashed line).}
	\label{fig:Zare}
\end{figure}

Scattering experiments were carried out for D$_2$ initial states built upon
$\ket{\psi_+}$, $\ket{\psi_-}$, which represent D$_2$ oriented at $\pm 45$ degrees relative to
the collisional flight axis, and the superposition $\ket{\psi_X} = 0.577 \left( \ket{\psi_+} - \ket{\psi_-}\right)$.
The measured differential cross-section $\sigma_X(\theta)$  from $\ket{\psi_X}$ did not correspond to the incoherent sum of the measured differential cross-sections $\frac{\sigma_+ + \sigma_-}{2}$ from $\ket{\psi_+}$ and $\ket{\psi_-}$, thereby demonstrating the role of interference. 
The deepened minimum arising from destructive interference
in the product differential cross section for rotational relaxation into the $j=0$ final state of D$_2$ , the central result of Ref.\cite{Zhou2021} is shown in our computational reproduction of their experimental results in Figure \ref{fig:Zare} (a). Note that theoretical simulations of these experiments have also been reported \cite{Jambrina2022}.

The three states above are, however, all linear combinations of the D$_2$ magnetic sublevels, which form
the  $m$-state basis: 
\begin{equation}
	\begin{split}
		\ket{\psi_\pm}=0.25 \ket{m=0}&\pm 0.612 [\ket{m=-1}-\ket{m=1}]\\&+0.306 [\ket{m=2}+\ket{m=-2}]
	\end{split}
\label{eq: decomp_pm}
\end{equation}
and 
\begin{equation}
	\ket{\psi_X}=\frac{1}{\sqrt{2}}\left(\ket{m=+1}-\ket{m=-1}\right).
\end{equation}
 In this basis there are no interference terms since the measured differential cross section involves integration over the azimuthal
angle $\phi$ and depends  exclusively on the polar angle $\theta$ \cite{Devolder2020}. Hence,
in the case where one colliding particle (e.g. D$_2$) is prepared in an m-superposition while the internal state of the other colliding particle (He) remains fixed, there can be no interference between these states, as each state within the superposition has a distinct internal projection\cite{Omiste2018,Devolder2020}. However, the differential cross section for the $j'=0$ 
state in the scattering of He + D$_2$, shown in Figure \ref{fig:Zare} (b), is exactly the same as that in Figure \ref{fig:Zare} (a), even if there are no interference terms! 
\textit{Hence, whether interference does or does
not contribute to the observed $\sigma(\theta)$ is clearly basis set dependent.}
The CCS matrix and various transformations associated with this experiment are discussed in the next section.

\subsection{Interferences---Evident or not}

To illustrate the appearance or absence of interference, consider the S-matrix elements 
that we obtained by fitting the measurements in these experiments \cite{Zhou2021}. The CCS matrices $\boldsymbol{\mathcal{C}^{half}}$ at two sample scattering angles, $\theta=\frac{\pi}{4}$ and $\frac{\pi}{2}$ and the final state of D$_2$(v=2, j'=0) 
in the m-state basis are: \begin{equation}
	\boldsymbol{\mathcal{C}}\left(\frac{\pi}{4}\right)= \begin{pmatrix}
		0.64 & 0& 0&0 &0 \\
		0&  3.56 & 0 & 0 & 0 \\
		0 & 0 &7.25 & 0&0\\
		0 & 0& 0 & 3.56 & 0\\
		0 & 0 & 0& 0& 0.64 
	\end{pmatrix}
	\label{eq:Zare_FSB1}
\end{equation}
and 
\begin{equation}
	\boldsymbol{\mathcal{C}}\left(\frac{\pi}{2}\right)= \begin{pmatrix}
		3.49 & 0& 0&0 &0 \\
		0&  0.0021 & 0 & 0 & 0 \\
		0 & 0 &3.88 & 0&0\\
		0 & 0& 0 & 0.0021 & 0\\
		0 & 0 & 0& 0&  3.49 
	\end{pmatrix}.
	\label{eq:Zare_FSB2}
\end{equation}
As noted above, these matrices are diagonal and hence display no interference structure,
as imposed by the symmetry \cite{Omiste2018}. In this case, the m-state basis corresponds to the eigenvector basis. This is in contrast with the example of $^{85}$Rb+ $^{85}$Rb scattering, where interference in the m-state basis survives after the integration over the angles because some states have the same total internal projection. Note that, as expected, the CCS matrices at the
two indicated sample angles differ from one another, since they arise directly from the scattering 
amplitudes (\ref{eq:CCSM}).

As evidenced experimentally, and noted above, scattering from the superposition  $\ket{\psi_X}=0.577*\left[\ket{\psi_+}-\ket{\psi_-}\right]$ in the $\ket{\psi_+}$ and  $\ket{\psi_-}$ basis states  does display interference. Note that since the $\ket{\psi_X}$ state is written as a superposition of only the $\ket{\psi_+}$ and  $\ket{\psi_-}$ states, the analysis of the experiments does not require other states from the basis that includes $\ket{\psi_+}$ and  $\ket{\psi_-}$  and which would span the same Hilbert space as the 5 m-states. Therefore, for this analysis, only a 2x2 CCS matrix is necessary. To obtain the CCS matrix in the subspace $|{\ket{\psi_+},\ket{\psi_-}}$, equation (\ref{eq:transf_non-orth}) is used with the matrix $\boldsymbol{P}$, obtained from the coefficients in the decomposition (Eq. (\ref{eq: decomp_pm})) of the $\ket{\psi_+}$ and the $\ket{\psi_-}$ states: 
\begin{equation}
	\boldsymbol{P}=\begin{pmatrix}
		0.306 & 0.612 &0.25 & 0.612 & 0.306 \\ 
		0.306 & -0.612 & 0.25 & -0.612 & 0.306  
	\end{pmatrix}
\end{equation}
and with the overlap matrix given by:
\begin{equation}
	\boldsymbol{S^{ovlp}}=
	\begin{pmatrix}
		1 &-0.4993\\
		-0.4993 &1
	\end{pmatrix}.
\end{equation}
The CCS matrices obtained at $\theta=\frac{\pi}{4}$ and $\frac{\pi}{2}$ in the subspace $\ket{\psi_+}$ and $\ket{\psi_-}$ are :
\begin{equation}
	\boldsymbol{\mathcal{C}'}\left(\frac{\pi}{4}\right)= \begin{pmatrix}
		6.13 &-5.84 \\
		-5.84 & 6.13 \\
	\end{pmatrix}
	\label{eq:Zare_basis1_+-}
\end{equation}
\begin{equation}
	\boldsymbol{\mathcal{C}'}\left(\frac{\pi}{2}\right)= \begin{pmatrix}
		0.23 & 0.22 \\
		0.22 & 0.23 \\
	\end{pmatrix}.
	\label{eq:Zare_basis2_+-}
\end{equation}
The off-diagonal elements of these CCS matrices are non-zero, showing the presence of interference observed in the experiments. The emergence of interference in this basis can thus be explained by two considerations. First, the $\ket{\psi_+}$ and $\ket{\psi_-}$ states mix the $m$-states having different values of cross sections resulting in a non-diagonal CCS matrix. Secondly, the non-orthogonality between $\ket{\psi_+}$ and $\ket{\psi_-}$ also induces the revival of interference by enhancing the path distinguishability. 

\subsection{Maximal Interference Basis}

The collisions between D$_2$ molecules and helium atoms can serve as an interesting illustration of the benefits of the maximal interference basis. The 5 dimensional quantum defined by Eq. (\ref{eq:QFT_transf}) Fourier transform is given by:
\begin{widetext}
\begin{equation}
	QFT=\frac{1}{\sqrt{5}}\begin{pmatrix}
		1 & 1 & 1 & 1 & 1\\
		1& e^{\frac{2\pi i}{5}} & e^{\frac{4\pi i}{5}} & e^{\frac{6\pi i}{5}} & e^{\frac{8\pi i}{5}} \\
		1 & e^{\frac{4\pi i}{5}} & e^{\frac{8\pi i}{5}} & e^{\frac{12\pi i}{5}} & e^{\frac{16\pi i}{5}}\\
		1 & e^{\frac{6\pi i}{5}} & e^{\frac{12\pi i}{5}} & e^{\frac{18\pi i}{5}} & e^{\frac{24\pi i}{5}}\\
		1 & e^{\frac{8\pi i}{5}} & e^{\frac{16\pi i}{5}} & e^{\frac{24\pi i}{5}} & e^{\frac{32\pi i}{5}}
	\end{pmatrix}.
\end{equation} 
By applying this transformation to the CCS matrices (Eqs. (\ref{eq:Zare_FSB1}) and (\ref{eq:Zare_FSB2})), the corresponding matrices in the Fourier basis are obtained:

\begin{equation}
	\boldsymbol{\mathcal{C}'_{Fourier}}\left(\frac{\pi}{4}\right)= \begin{pmatrix}
		3.13 & -1.36-0.99i& 0.12+0.36i&0.12-0.36i &-1.36+0.99i \\
		-1.36+0.99i&  3.13 & -1.36-0.99i & 0.12 +0.36i & 0.12-0.36i \\
		0.12-0.36i & -1.36+0.99i &3.13 & -1.36-0.99i&0.12+0.36i\\
		0.12+0.36i & 0.12-0.36i& -1.36+0.99i & 3.13 & -1.36-0.99i\\
		-1.36-0.99i & 0.12+0.36i & 0.12-0.36i& -1.36+0.99i& 3.13 
	\end{pmatrix}
	\label{eq:Zare_int_FSB1}
\end{equation}
and 
\begin{equation}
	\boldsymbol{\mathcal{C}'_{Fourier}}\left(\frac{\pi}{2}\right)= \begin{pmatrix}
		2.1728 & 0.29 +0.21i& 0.37+1.15i&0.37-1.15i &0.29-0.21i \\
		0.29-0.21i&  2.1728 & 0.29+0.21i & 0.37+1.15i & 0.37-1.15i \\
		0.37-1.15i &0.29-0.21i  &2.1728 & 0.29+0.21i&0.37+1.15i\\
		0.37+1.15i & 0.37-1.15i& 0.29-0.21i & 2.1728 & 0.29+0.21i\\
		0.29+0.21i & 0.37+1.15i & 0.37-1.15i& 0.29-0.21i& 2.1728
	\end{pmatrix}.
	\label{eq:Zare_int_FSB2}
\end{equation}
\end{widetext}
All the states interfere with one another and give the same cross section. The control index $R_c$ is 0.97 (i.e an almost complete control), as opposed to 0 in the m-state basis (i.e. no coherent control). In the m-state basis (corresponding to the eigenvector basis in this case), the range of variation from the minimum value to the maximum value could only be achieved by varying the relative populations. In contrast, in the Fourier basis, this variation is accomplished only by modifying the relative phases. Thus, in the CCS eigenvector basis the
interference terms play no role, whereas in the Fourier basis the interference
terms play the maximum possible role. In all other bases,  there are both classical and quantum interference contributions to control.

	\section{Conclusion}

We emphasize that 
interference is basis-dependent and show that the CCS matrix provides a valuable tool for examining this dependence in scattering problems. It can be generalized to arbitrary quantum processes, a subject of future work. Bases, in which the CCS matrix is diagonal do not allow interference whereas a CCS matrix
with off-diagonal elements implies the possibility of interference. Transformation operations $\boldsymbol{U}$  between bases can alter their interference structure. For example, the non-commutativity of $\boldsymbol{U}$ and the CCS matrix, as well as the non-orthogonalilty of one of the bases can readily introduce an interference 
structure.  There are two extreme bases: (a) the CCS eigenbasis with zero interference and hence maximal path distinguishability, and (b) the Fourier basis of the eigenvectors with maximal interference and
hence zero path distinguishability.  It is crucial to note that building
such states in the laboratory may imply states that are experimentally challenging to construct and hence
primarily of academic interest.

 Interference issues are of particular interest to coherent control, a methodology for controlling 
molecular processes through interference discussed here in detail. In this case care must be taken
to distinguish between control of a cross section, and control of a branching ratio into different
product states. In general it is the latter that is of greater interest, Hence, although it is 
always possible, in the case of orthogonal bases, to transform to a basis where there is no interference, in a cross section, this is not the case for the branching ratio. 

Finally, we note that the recent experiment \cite{Zhou2021} on the inelastic differential
scattering cross section in cold He + D$_2$ collisions provides an excellent example of 
the basis sect dependence of interference contributions. Although it is possible
to obtain the observed $\sigma(\theta)$ without invoking interference, the
fact that interference can be seen in one basis implies that the scattering does
preserve interference and phase effects, as expected since the S-matrix constitutes a unitary transformation. Consideration of the basis set dependence of the role of interference in other
(non-scattering) phenomena, is the subject of future efforts.

{\bf Acknowledgements:} Issues of the type discussed here were brought to the attention of PB
by Dr. Johannes Floss several years ago, but not developed. We are grateful to him for initial discussions.
This work was supported by the U.S. Office of Scientific Research (AFOSR) under contract number
FA9550-22-1-0361.

 \bibliography{FSB}

\end{document}